\documentclass[manuscript]{aastex}

\slugcomment{The Astronomical Journal}

\shorttitle{Active Asteroid 311P}
\shortauthors{Jewitt et al.}

\begin{document}


\title{The Nucleus of  Active Asteroid 311P/(2013 P5) PANSTARRS}


%\author{David Jewitt\altaffilmark{1,2}}
%\affil{$^1$ Department Earth, Planetary and Space Sciences,  and Department Physics and Astronomy, University of California,
%    Los Angeles, CA 90095 }
%
%%\author{C. D. Biemesderfer\altaffilmark{4,5}}
%%\affil{National Optical Astronomy Observatories, Tucson, AZ 85719}
%\email{jewitt@ucla.edu}

\author{David Jewitt$^{1,2}$, Harold Weaver$^3$,  Max Mutchler$^4$, Jing Li$^{1}$, Jessica Agarwal$^5$  and Stephen Larson$^6$
}
\affil{$^1$Department of Earth, Planetary and Space Sciences,\\
UCLA, 
595 Charles Young Drive East, 
Los Angeles, CA 90095-1567\\
$^2$Department~of Physics and Astronomy, \\
UCLA, 
430 Portola Plaza, Box 951547,
Los Angeles, CA 90095-1547\\
$^3$ The Johns Hopkins University Applied Physics Laboratory, \\
11100 Johns Hopkins Road, Laurel, Maryland 20723  \\
$^4$ Space Telescope Science Institute,\\
 3700 San Martin Drive, Baltimore, MD 21218 \\
$^5$ Max Planck Institute for Solar System Research, \\Justus-von-Liebig-Weg 3, 37077 G\"ottingen, Germany\\
$^6$ Lunar and Planetary Laboratory, \\
University of Arizona, 1629 E. University Blvd.
Tucson AZ 85721-0092 \\
}
\email{jewitt@ucla.edu}
%\and

%\author{R. J. Hanisch\altaffilmark{5}}
%\affil{Space Telescope Science Institute, Baltimore, MD 21218}


%\altaffiltext{1}{Visiting Astronomer, Cerro Tololo Inter-American Observatory.
%CTIO is operated by AURA, Inc.\ under contract to the National Science
%\altaffiltext{2}{Society of Fellows, Harvard University.}
%\altaffiltext{3}{present address: Center for Astrophysics,
%    60 Garden Street, Cambridge, MA 02138}
%\altaffiltext{4}{Visiting Programmer, Space Telescope Science Institute}
%\altaffiltext{5}{Patron, Alonso's Bar and Grill}


\begin{abstract}
The unique inner-belt asteroid 311P/PANSTARRS (formerly P/2013 P5) is notable for its sporadic, comet-like ejection of dust in nine distinct epochs spread over $\sim$250 days in 2013.  This curious behavior has been interpreted as the product of localized, equator-ward landsliding from the surface of an asteroid rotating at the brink of  instability.   We obtained new Hubble Space Telescope observations  to  directly measure  the nucleus and to search for evidence of its rapid rotation. We find a nucleus with mid-light absolute magnitude $H_V$ = 19.14$\pm$0.02, corresponding to an equal-area circle with radius 190$\pm$30 m (assuming geometric albedo $p_V$ = 0.29).   However, instead of providing  photometric evidence for rapid nucleus rotation, our data   set a lower limit to the lightcurve period, $P \ge$ 5.4 hour.  The dominant feature of the lightcurve is a V-shaped minimum,  $\sim$0.3 magnitudes deep, that is suggestive of an eclipsing binary.  Under this interpretation, the time-series data are consistent with a  secondary/primary mass ratio, $m_s/m_p \sim$ 1:6,  a ratio of separation/primary radius, $r/r_p \sim$ 4 and an orbit period $\sim$0.8 days.  These properties lie within the range of  other asteroid binaries that are thought to be formed by rotational breakup.  While the lightcurve period is long, centripetal dust ejection is still possible if one or both components rotates rapidly ($\lesssim$ 2 hour) and has a small lightcurve variation because of azimuthal symmetry.  Indeed, radar observations of asteroids in critical rotation  reveal  ``muffin-shaped'' morphologies which are closely azimuthally symmetric and which show minimal lightcurves.   Our data are consistent with 311P being a close binary in which one or both components rotates near the centripetal limit.  The mass loss in 2013 suggests that breakup occurred recently and could even be on-going.    A search for  fragments that might have been recently ejected beyond the Hill sphere reveals none larger than effective radius $r_e \sim$ 10 m.
\end{abstract}




%% Note that for sources with brackets in their names, e.g.~[WEG2004] 14h-090,

\section{Introduction}

\begin{figure}
\epsscale{1.0}
%\plotone{prettypic.pdf}
\plotone{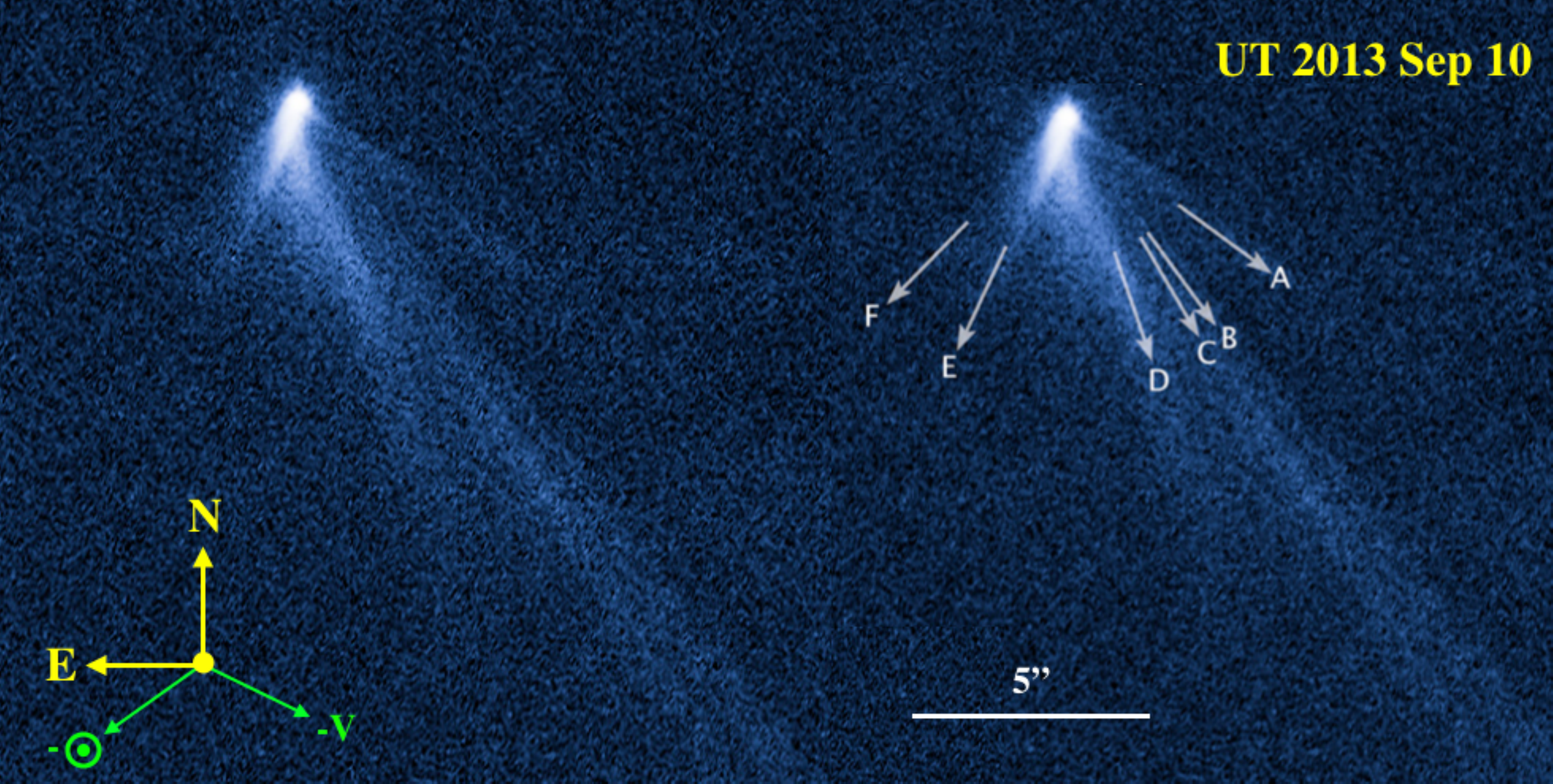}
\caption{311P imaged in an active state on UT 2013 September 10 (Jewitt et al.~2013).  Letters mark individual dust tails.  Three additional tails (G - I) were ejected after this image was taken (Jewitt et al.~2015b). The cardinal directions are indicated by yellow arrows while the projected antisolar vector (-$\odot$) and the negative heliocentric velocity vector (-$V$) are shown in green.  \label{prettypic}}
\end{figure}

Active asteroids occupy the orbits of asteroids but show the physical appearances of comets, caused by transient mass loss.  Their properties indicate  a surprising diversity of mass-loss processes, including impact, thermal fracture and suspected rotational break-up (Jewitt et al.~2015a).
One of the strangest such objects is 311P/PANSTARRS (formerly P/2013 P5, hereafter ``311P'').  
The orbit has semimajor axis, $a$ =  2.189 AU, eccentricity $e$ = 0.115, and inclination $i$ = 5.0\degr, leading to an asteroid-like Tisserand parameter measured relative to Jupiter, $T_J$ = 3.662.  311P orbits near the inner edge of the asteroid belt, where most objects are thought to be highly metamorphized S-types, with suspected meteorite counterparts in the LL chondrites (Keil 2000, Dunn et al.~2013).  In addition, the orbital elements of the 10$^9$ year old Flora asteroid family (Dykhuis et al.~2014) are similar to those of 311P, suggesting that it might be a member.  Despite its inner-belt orbit and likely refractory nature, 311P displayed a unique, multi-tailed morphology (Figure \ref{prettypic}) caused by episodic dust ejection events (Jewitt et al.~2013, 2015b) spanning the period UT 2013 April 15 (pre-perihelion heliocentric distance $r_H$ = 2.304 AU) to December 26 (post-perihelion $r_H$ = 1.989 AU).  Each tail is a ``synchrone'' formed by the action of radiation pressure on $\sim$10$^5$ kg of dust launched with initial speeds $<$1 m s$^{-1}$.   The tails are separated by intervals of inactivity lasting from weeks to months.   Jewitt et al.~(2013, 2015b) conjectured that 311P might be rotating near the brink of centripetal instability, allowing small avalanches of debris to leave from the equatorial regions and then swept into the observed discrete tails by radiation pressure.  This model was quantified and described as ``mass-shedding''  by  Hirabayashi et al.~(2015) and Scheeres (2015).  The importance of rotation in asteroidal mass loss was soon reinforced by observations of another active asteroid, P/2013 R3, in which the body of the asteroid has split into a dozen or more components (Jewitt et al.~2014, 2017). 

In this paper we present new Hubble Space Telescope (HST) observations taken to further examine the development of activity in 311P and to better characterize the nucleus.  A particular objective is to seek evidence for rapid rotation, expected if the mass-shedding instability hypothesis is correct.

\section{Hubble Space Telescope Observations}

Observations using the HST were taken in a series of seven epochs using the Wide Field Camera 3 (WFC3)  in imaging mode, allocated under program GO 13866.  The UVIS channel of the WFC3 camera houses two charge-coupled devices each having 2051$\times$4096 pixels, with square pixels 0.04\arcsec~on a side. The field of view is 162\arcsec$\times$162\arcsec.   The image scale projected to the distance of 311P varied by a factor of two from 38 km pixel$^{-1}$ in UT 2015 March to 76 km pixel$^{-1}$ in UT 2015 July.      We used the F350LP filter, which has a transmission full width at half-maximum (FWHM) = 4758\AA~and an effective central wavelength 6230\AA~when used to observe a solar-type source.  

\begin{figure}
\epsscale{1.0}
%\plotone{obs_plot.pdf}
\plotone{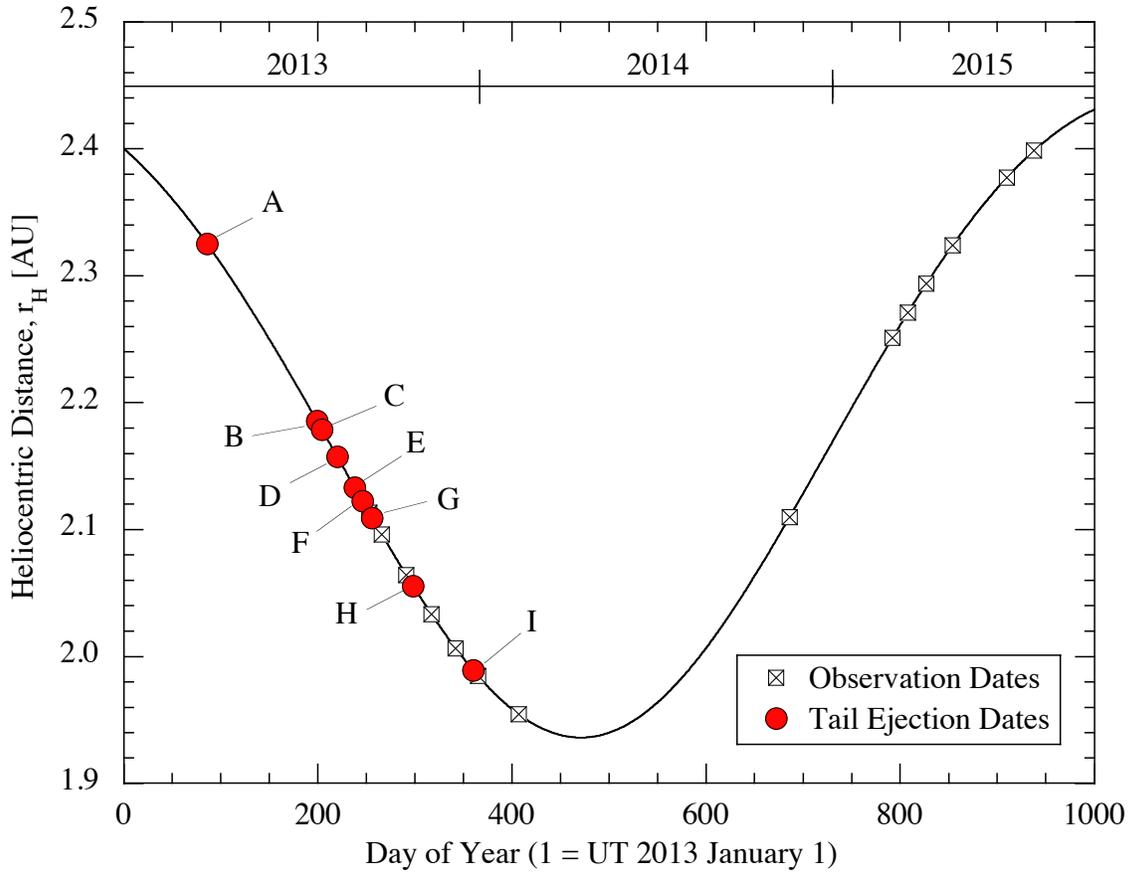}
\caption{Heliocentric distance (AU) vs. the dates of observation (shown as $\boxtimes$) and of dust tail ejections (red filled circles).  The dates are expressed as Day of Year for convenience, where DOY = 1 on UT 2013 January 1. The tail ejections are labeled A - I (compare with Figure \ref{prettypic}), as in Jewitt et al.~(2015).  \label{obs_plot}}
\end{figure}

Five images of 400 s to 420 s duration were taken in each orbit of HST, all significantly contaminated by cosmic rays. We combined the five dithered images using a median filter to reject cosmic rays and detector artifacts, after shifting to align the images on 311P. Magnitudes determined from the orbital median images were measured using a set of concentric photometry apertures.  Given the stability of the HST PSF, and with no coma or tail structures evident in the data, we chose to employ very small photometry apertures (0.2\arcsec~and 1.0\arcsec~radius) in order to minimize uncertainties due to the sky background. We obtained sky subtraction from a concentric annulus having inner and outer radii 1.0\arcsec~and 4.0\arcsec, respectively.  The photometric calibration assumes that a V = 0 solar-type source would give a count rate in the F350LP filter of 4.72$\times$10$^{10}$ s$^{-1}$, as estimated from the on-line exposure time calculator.  The average magnitudes within the 0.2\arcsec~aperture in each HST visit are listed in Table (\ref{photometry}).  The distribution of the HST observations is shown in Figure (\ref{obs_plot}), where we have also marked and labelled (with red circles) the inferred dates of dust ejections from Jewitt et al.~(2013, 2015b).

To measure the magnitudes within individual images we proceeded as follows.  We subtracted the orbital median image from the individual images within the same orbit to remove as far as possible any real signals in the data, leaving only the variable component of 311P, cosmic rays and noise.  Then, we removed cosmic rays in the vicinity of the nucleus by hand, replacing them with the average signal from adjacent pixels.  Finally, we added back the median image to obtain five cosmic-ray free images of 311P per HST orbit.  Photometry of these cleaned images was performed as described above.  In a few cases, we found images struck by cosmic rays so close to the nucleus of 311P that no removal was feasible.  In these cases, we simply omit the data from further consideration.

\clearpage

\section{Measured Properties}
\subsection{Activity}
\label{Activity}
\begin{figure}
\epsscale{.70}
%\plotone{V_vs_DOY.pdf}
\plotone{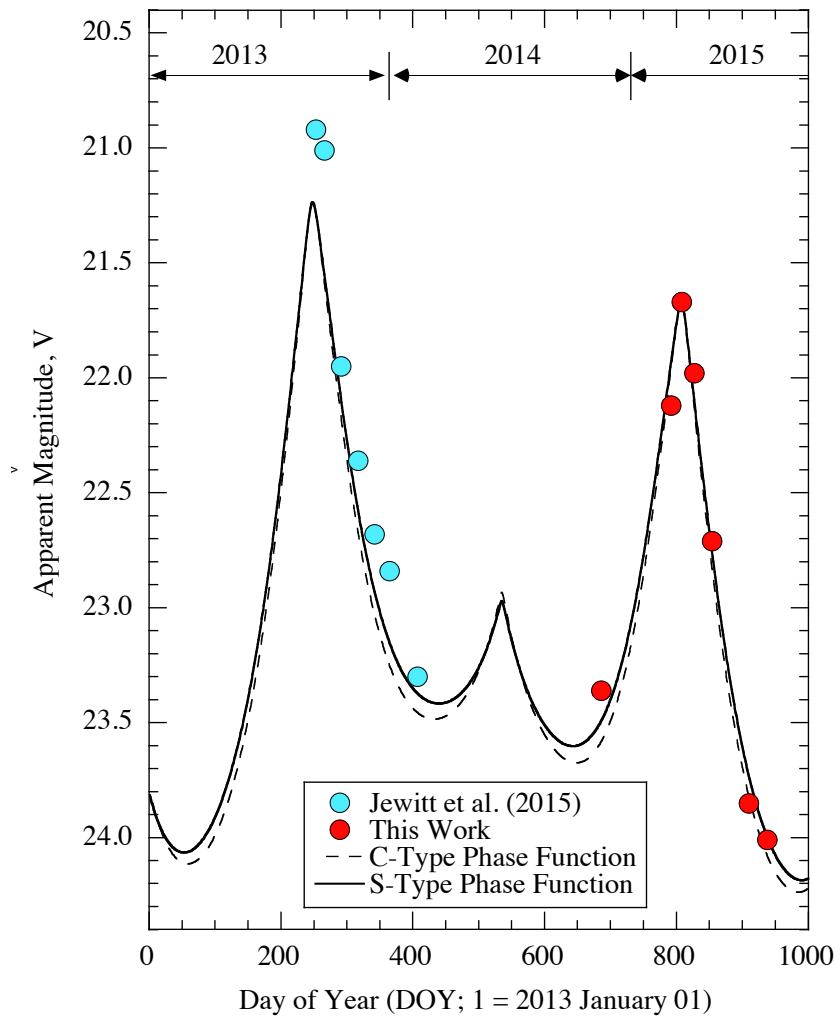}
\caption{Apparent magnitude measured within a 0.2\arcsec~radius aperture as a function of time, measured in days, with DOY = 1 on 2013 January 01.  Blue and red symbols distinguish  measurements from Jewitt et al.~(2015) from those from Table (\ref{photometry}), respectively.  Photometric error bars are smaller than the plot symbols. Solid and dashed lines show the model brightnesses for a spherical body following S-type and C-type phase functions, respectively, as described in the text. \label{V_vs_DOY}}
\end{figure}

The prominent dust tails evident in earlier observations of 311P (Jewitt et al.~2013, 2015b) were absent in  the new observations presented here.  Figure (\ref{V_vs_DOY}) shows the apparent magnitude for both old and new data as a function of time, expressed as Day of Year (DOY), where DOY = 1 is defined as UT 2013 January 1.  The symbols in the figure represent the averages of measurements taken within a given HST orbit.  The associated photometric uncertainties, expressed as the 1$\sigma$ errors on the means, are comparable to or smaller than the sizes of the plot symbols.  Also plotted on the figure are curves showing the expected time-variation of the magnitude computed from 

\begin{equation}
V = H + 2.5\log_{10}\left(r_H^2 \Delta^2\right) -  2.5\log_{10}(\Phi(\alpha))
\label{abs}
\end{equation}

\noindent where $r_H$ and $\Delta$ are the heliocentric and geocentric distances expressed in AU and $\Phi(\alpha) \le 1$ is the phase function at phase angle $\alpha$.   The absolute magnitude, $H$, defines the normalization of the curves to the data.  It is the apparent magnitude that would be observed if the object were to be located at $r_H = \Delta$ = 1 AU and $\alpha$ = 0\degr.  The phase function of 311P is unmeasured but, over the range of phase angles at which 311P was observed, is unlikely to have a major effect on the interpretation of the photometry.   To show this, we plot in Figure (\ref{V_vs_DOY}) two phase functions representing the nominal behavior of C-type (dashed line) and S-type (solid line)  asteroids (characterised by assuming parameter $G$ = 0.15 and 0.25, respectively, in the formulation by Bowell et al.~1989).  The figure shows that variations in $r_H$ and $\Delta$ dominate the long-term photometric variations of 311P and that the effect of phase angle, $\alpha$, is comparatively modest at the phase angles sampled in our data.  In the rest of this work, we assume the phase function of S-type asteroids, guided by the optical color of 311P (Hainaut et al.~2014) and by the location of this object in the inner region of the asteroid belt, where S-types are most abundant.

\begin{figure}
\epsscale{.80}
%\plotone{HV_vs_DOY.pdf}
\plotone{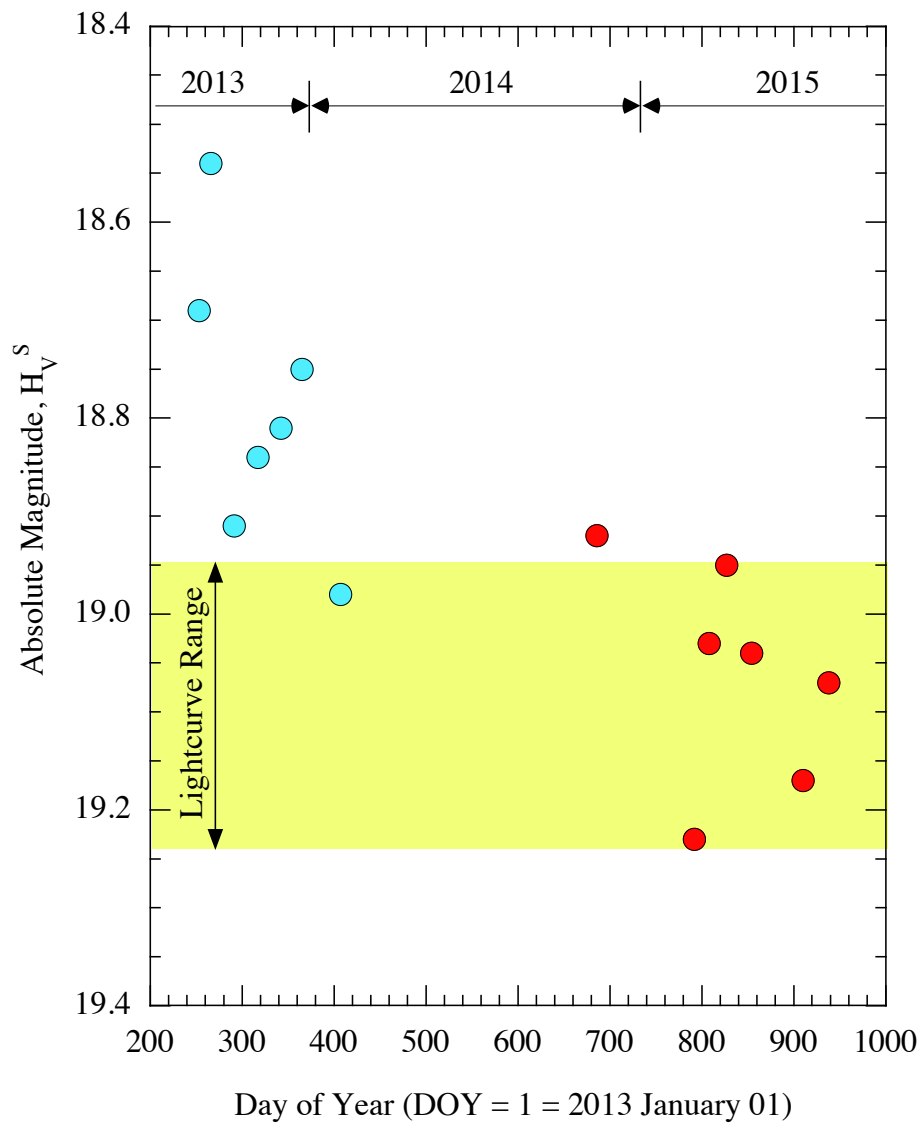}
\caption{Absolute magnitude measured within a 0.2\arcsec~radius aperture as a function of time, measured in days, with DOY = 1 on 2013 January 01.  Red and blue symbols distinguish the new measurements from this paper from measurements published in Jewitt et al.~(2015), respectively.  An S-type phase function has been assumed.  The yellow shaded region marks our best estimate of the nucleus lightcurve range.  \label{HV_vs_DOY} }
\end{figure}

By fitting the model curves to the data, we observe that most of the early measurements (specifically those from discovery at DOY = 230 to DOY $\sim$ 400) imply an $H$ systematically brighter than later data  (Figure \ref{V_vs_DOY}).  This is clearly seen in Figure (\ref{HV_vs_DOY}), where we plot the absolute magnitude as a function of time.  We  attribute this fact to early dust activity in 311P and we use only the 2015 photometry from Table (\ref{photometry}) to obtain our best estimate of $H$ for the nucleus of 311P.   The mid-light value, $H = 19.14\pm0.02$,  is consistent with a limit, $H \ge 18.98\pm0.10$, placed from an analysis of data taken in 311P's active phase (Jewitt et al.~2015b).  Short-term variations about the mean, in the range $18.95 \le H \le 19.23$,  are likely due to lightcurve effects.   

The absolute magnitude and the scattering cross-section, $C_e$, are related by 

\begin{equation}
p_V  C_e = 2.25\times10^{22} \pi  10^{-0.4(H - m_{\odot})}
\label{inversesq}
\end{equation} 

\noindent where $m_{\odot}$ = -26.75 is the apparent V magnitude of the Sun (Drilling and Landolt 2000).   We set $p_V$ = 0.29$\pm$0.09, which is the average geometric albedo of the Flora family asteroids as reported by Masiero et al.~(2013), to find that $C_e$ varies from $C_{min}$ = 0.10 km$^2$ to $C_{max}$ = 0.13 km$^2$, with a mid-light value $C_e = 0.11\pm0.04$ km$^2$, where the error is dominated by the uncertainty on $p_V$.  The mid-light cross-section corresponds to the area of a circle of effective radius $r_e = (C_e/\pi)^{1/2}$ = 190$\pm$30 m (again, the uncertainty on the effective radius is dominated by the uncertainty on the albedo).  This is consistent with an estimate based on earlier HST photometry ($r_e \le 200 \pm 20$m, Jewitt et al.~2015b) but considerably  larger than the radius 15 $\le r_e \le$ 65 m estimated by Moreno et al.~(2014) based on a model of the motion of dust.  Note that, if our albedo assumption is wrong, both the effective radius and the disagreement with the estimate by Moreno et al.~(2014) would likely be larger (e.g.~if $p_V$ = 0.04, as for a C-type object, the inferred radius would be larger by the ratio (0.29/0.04)$^{1/2}$ = 2.7.  %(They inferred the nucleus size by assuming that  the minimum dust velocity  corresponds to the gravitational escape velocity from the non-rotating nucleus. Their underestimate of the nucleus size may be an independent indication that the effective escape velocity is reduced by rapid rotation.)

We sought spatially resolved evidence for dust in the data of Table (\ref{photometry}) by comparing photometry within concentric apertures.  Specifically, we used $\delta V = V_{0.2} - V_{0.4}$ to measure the excess contribution from dust in the 0.2\arcsec~to 0.4\arcsec~radius range.  Light in this annulus includes contributions from the wings of the point-spread function, together with additional light scattered by near-nucleus dust.  In all but the UT 2014 November 17 images, the value of $\delta V$ is consistent with expectations based on modeling the point-spread function of HST, for which we used the TinyTim software.  Considering all the data, we find a median $\delta V_m = 0.092$ magnitudes.  On UT 2014 November 17, we find median $\delta V = 0.15$ magnitudes and interpret the difference, $\delta V - \delta V_m = 0.06$ magnitudes, as caused by light scattered from near-nucleus dust, with an average surface brightness in the 0.2\arcsec~to 0.4\arcsec~annulus of $\Sigma(0.2\arcsec-0.4\arcsec)$ = 25.5 magnitudes arcsec$^{-2}$.  

The relation between the surface brightness of a steady-state dust coma measured at radius $\theta$\arcsec~and the integrated brightness, $V_{Tot}$, measured within an aperture is (Jewitt and Danielson 1984)

\begin{equation}
V_{Tot} = \Sigma(\theta) - 2.5\log_{10}\left(2 \pi \theta^2\right).
\label{JD84}
\end{equation}

Substituting into Equation (\ref{JD84}) we find coma magnitude $V_{Tot}$ = 26.12, compared with the measured magnitude of 311P on UT 2014 November 17 of $V$ = 23.47$\pm$0.01 (Table \ref{photometry}).  If the model coma brightness is subtracted from the measured brightness, we find a bare nucleus magnitude $V$ = 23.57, about 0.1 magnitudes fainter than in the Table.  The exact value of the coma correction is uncertain because, for instance, we do not know that the coma is in steady state, as assumed.  Steeper (flatter) surface brightness profiles would lead to larger (smaller) coma contributions to the photometry.  Nevertheless, it appears that $\sim$10\% of the signal measured on UT 2014 November 17 results from dust while, in all other epochs in Table (\ref{photometry}), we find no evidence for coma. 

Large particles tend to be ejected slowly and their motions are confined to follow along the projected orbit, forming low surface brightness, nearly parallel-sided ``trails'' (e.g.~Kim et al.~2017).  We find no evidence for such a large-particle trail, even in our deepest imaging data (Figure \ref{march19}).  Particles released from the nucleus in 2013 with zero initial velocity would be pushed by radiation pressure to an angular distance $\sim$10\arcsec~in our 2015 data if their sizes were $\lesssim$1 cm.  The apparent absence of such particles is consistent with the cross-section weighted mean particle size (3.4 mm) inferred in the tails in earlier images (Jewitt et al., 2015).  Particles of this size would have been dispersed by radiation pressure in the two years since the 2013 outbursts.  The low surface brightness excess in the 0.2\arcsec~to 0.4\arcsec~annulus, described above, is most likely an indicator of small dust grains expelled by late-stage activity.

\subsection{Search for Secondary Objects}
\label{search}

A $\sim$190 m sized body has a very small region of gravitational control, given by the Hill radius,

\begin{equation}
R_H = a  \left(\frac{r_e}{r_{\odot}}\right) \left(\frac{\rho}{3\rho_{\odot}}\right)^{1/3}
\label{Hill}
\end{equation}

\noindent where $a$ is the semimajor axis, $r_{\odot}$ and $\rho_{\odot}$ are the radius and the density of the Sun, respectively, and $r_e$ and $\rho$ are again the effective radius and the density of 311P.  
The latter is unmeasured and so we adopt the mean density of small S-type asteroids, $\rho_S$ = 2700$\pm$500 kg m$^{-3}$ as determined by Carry (2012).  For comparison, the LL chondrite meteorites are thought to originate from the inner asteroid belt, in the orbital vicinity of 311P.  Their average density has been measured at $\rho$ = 3200$\pm$200 kg m$^{-3}$ (Consolmagno et al.~2008).  The (small) density difference likely results from the existence of macroporosity in the measured asteroids and from a selection effect whereby denser, stronger asteroid fragments better survive passage through the Earth's atmosphere than do less dense, weaker fragments. We assume that the density of 311P is well represented by $\rho_S$.  

Substituting orbital semimajor axis $a$ = 2.189 AU, $\rho$ = 2700 kg m$^{-3}$, $\rho_{\odot}$ = 1400 kg m$^{-3}$, $r_e$ = 190 m and $r_{\odot}$ = 7$\times$10$^8$ m, we find $R_H$ = 80 km.  When observed from minimum distance $\Delta \sim$ 1.3 AU (as in 2015 March, see Table \ref{geometry}), the Hill radius subtends an angle $\theta_H = R_H/\Delta \sim$ 0.08\arcsec, comparable to the $\sim$0.08\arcsec~resolution corresponding to Nyquist (2 pixel) sampling of the WFC3 images.  Therefore, any existing gravitationally bound companions to 311P fall  beneath the resolution of the HST data.  

In some models of rotational breakup (e.g.~Jacobson and Scheeres 2011, Walsh et al.~2012, Boldrin et al.~2016, discussed later) temporarily stable companion objects are launched from orbit around the primary by complex gravitational interactions, eventually creating asteroid pairs.  With  radius $r_e$ = 190 m and density $\rho$ = 2700 kg m$^{-3}$, a non-rotating and spherical 311P would have a gravitational escape speed $V_e \sim$ 0.23 m s$^{-1}$.  If traveling at $V_e$, an escaping body would spend 4$\times$10$^6$ s (48 days) within 1\arcsec~of the primary nucleus and $\sim$2$\times$10$^8$ s (7 years) within 50\arcsec~of it.   Objects ejected in 2013, $\sim$400 days before the present observations, would have traveled $\sim$10$^7$ m (10\arcsec) from the nucleus at speed $V_e$ and would be well within the WFC3 field of view.  Therefore, although we cannot hope to resolve bound companions inside the Hill sphere, it remains worthwhile to search the HST images for evidence of unbound objects that might be slowly leaving the vicinity of the primary.  

To this end, we examined the WFC3 images from UT 2015 March 19 (when $\Delta$ = 1.283 AU) in search of co-moving field objects near 311P. We first aligned the images on 311P and then formed image composites using different combinations of the 20 separate images.  Each composite image consisted of 10 separate 400 s exposures, for an effective exposure of 4000 s.  Pairs of image composites were then visually compared in order to distinguish spurious sources formed by chance noise clumps and overlapping background trail residuals from potentially real ones.  Because of the large parallax motions in these data, background stars and galaxies were not a significant source of confusion except to the extent that they contribute to the background noise.  Numerous cosmic ray tracks in the CCD images were successfully removed in the computation of the composites.  

\begin{figure}
\epsscale{.90}
%\plotone{stars.pdf}
\plotone{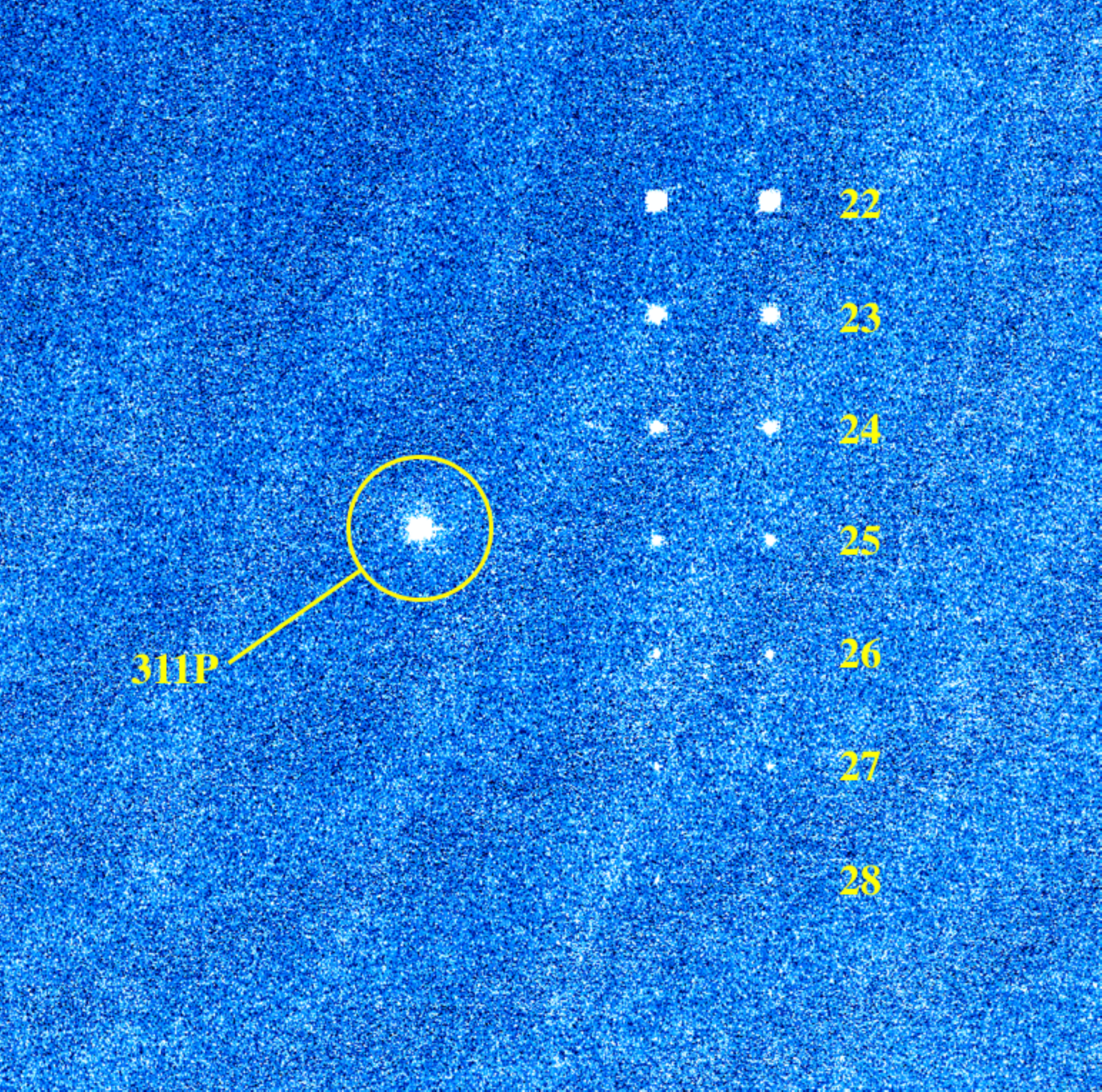}
\caption{Composite 4000 s image from UT 2015 Mar 19 showing 311P (circled) and digitally added stars having apparent $V$ magnitudes from 22 to 28, as labeled. Diagonal streaks in the image are the residuals left by parallactically trailed field galaxies.  Width of image shown is 40\arcsec.  \label{stars}}
\end{figure}

No co-moving objects were found in a region (limited by the position of 311P on the WFC3 CCD) extending $\sim$40\arcsec~to the south, 120\arcsec~to the north, and $\sim$80\arcsec~east and west of the nucleus. The limiting magnitude of the image composites was estimated in two ways.  First, we  digitally added sources of known brightness to the data and searched for them in the same way as we searched for real companions.   An example grid of digital stars is shown in Figure (\ref{stars}) labeled by their $V$ magnitudes. It may be seen that the visibility of the faintest sources  is limited by small variations in ``sky'' noise caused by the residual images of trailed galaxies.  Using the digital stars, we estimate an effective limiting magnitude $V$ = 28.3, applicable at distances from 311P greater than $\sim$1\arcsec.  At smaller separations, the surface brightness from the wings of the point spread function reduces the sensitivity.  Separately, we used the on-line exposure time calculator for the WFC3 camera (\url{http://etc.stsci.edu/})  to obtain an independent estimate of the limiting magnitude.  For 10 exposures of 400 s each, on a  source with G2V spectral type observed using the  350LP filter, a 3$\sigma$ detection is expected at $V$ = 27.8.  This is slightly poorer than found empirically, a fact which we attribute to our ability to detect objects with formal significance $<$3$\sigma$ in blinked-pair data.  However, in order to be conservative, we take $V$ = 27.8 as the upper limit to the allowable brightness of any point-like, co-moving companion.

By Equation (\ref{inversesq}), $V$ = 27.8 corresponds to a spherical effective radius $r_e$ = 11 m (again, assuming $p_V$ = 0.29) and sets a practical upper limit to the size of any objects ejected from 311P and still remaining within the vicinity of the nucleus.  Such a body would have a mass $\sim$2$\times$10$^{-4}$ times that of the central body.

\subsection{Photometric Variations}
Figure (\ref{composite}) shows the individual  lightcurves from the single-orbit visits to 311P.  Each visit is limited to  $\lesssim$40 minutes duration by the orbital motion of HST, but even during this short time the apparent brightness of 311P is observed to vary by an amount large compared to the $\pm$0.01 magnitude uncertainty of measurement.  For example, on UT 2015 March 3 and June 29, the brightness changes by $\sim$0.2 magnitudes, or 20$\times$ the measurement uncertainty.  The plotted photometry is extracted from a 0.2\arcsec~radius photometry aperture, corresponding to $\sim$190 km at $\Delta$ = 1.3 AU.  Within this aperture the signal is dominated by light from the nucleus and so we presume that most or all of the measured photometric variation is due to rotational modulation of the scattered light, owing to the aspherical shape of the nucleus.  

Unfortunately,  the majority of the lightcurve segments in the Figure are so widely separated in time that it is not possible to combine them in order to reconstruct a unique rotational lightcurve.   Instead, the strongest lightcurve constraint is provided by the data from four consecutive orbits on UT 2015 March 19.  These four orbits span a period of 5.4 hours (Figure \ref{march19}) and, since the features of the lightcurve do not repeat, we conclude  that a limit to the  lightcurve period (which is not necessarily equal to the rotation period) may be set at $P_{\ell} \gtrsim$ 5.4 hours.  

\clearpage 

\section{Discussion}
We briefly consider possible models of the nucleus which conform to this period constraint while providing the centripetal instability inferred to drive intermittent dust loss.

\subsection{Albedo Variations}
If the brightness variations on 311P are due to azimuthal albedo variations, then the rotational period, $P$, and the lightcurve period are equal, $P = P_{\ell}$.  However, azimuthal albedo variations on most small bodies are modest to the point of being undetectable.  While we have no specific evidence in the case of 311P, we consider it unlikely that the lightcurve in Figure (\ref{march19}) is caused by surface albedo variations.

\subsection{Prolate Body}
The lightcurves of most small bodies in the solar system reflect variations in the projected cross-section caused by rotation.  Such geometrically produced lightcurves are doubly periodic (two maxima and two minima per rotation) meaning that $P = 2P_{\ell}$.   In this interpretation, the rotation period of 311P would be constrained by the data to be $P \ge$ 10.8 hr.  

The critical period at which a strengthless body becomes rotationally unstable depends on both its density and its shape.  Consider a prolate (American football shaped) body with a long axis of length, $2b$ and two, equal short axes of lengths $2a$. In rotation about one of the short axes, the gravity at the tips  equals the centripetal acceleration there at critical period

\begin{equation}
P_c = \left(\frac{3\pi}{G\rho}\right)^{1/2} \left(\frac{b}{a}\right).
\label{prolate}
\end{equation}

\noindent The axis ratio, $b/a$, can be estimated from the measured rotational brightness variations, $\Delta V$ using $b/a$ = 10$^{0.4\Delta V}$.  Substituting  $\Delta V$ = 0.3 magnitudes (Figure \ref{march19}), we find $b/a$ = 1.3 and this is formally a lower limit to the axis ratio because of the effects of projection (i.e.~the rotation axis might not be perpendicular to the line of sight).

Substituting $\rho$ = 2700 kg m$^{-3}$, $b/a$ = 1.3 into Equation (\ref{prolate}) we find $P_c = $ 2.6 hr, which matches the ``spin barrier'' detected in the rotations of most asteroids larger than $\sim$150 m in size, C- and S-type alike (e.g.~Carbognani 2017, and references therein). Given that  $P \gg P_c$, we conclude that 311P, interpreted as a single, rotating prolate body, is centripetally stable.  Forcing $P_c = 10.8$ hr in Equation (\ref{prolate}) gives $\rho <$ 160 kg m$^{-3}$ for centripetal instability, which we consider implausibly small.  The  lightcurve (Figure \ref{march19}) is thus inconsistent  with  centripetal instability models of the activity in 311P  (Jewitt et al.~2013, 2015b, Hirabayashi et al.~2015) if the nucleus is prolate.  

\subsection{Oblate Body}

Radar observations (e.g.~Ostro et al.~2006, Busch et al.~2011, Naidu et al.~2015) show that many rapidly rotating objects possess an oblate, discus or muffin-like, body-shape likely produced by equatorward migration of material from higher latitudes.  A perfectly oblate body in rotation about its minor axis is rotationally symmetric, would have no rotational lightcurve and so could not be identified using time-series photometry.  While  radar-observed asteroids are not perfectly oblate, the typical rotational variation in the cross-section is just a few percent (e.g.~Pravec et al.~2006), again challenging detection in our data.  We cannot appeal to such a body to explain the large photometric variations in 311P.  Specifically, the 0.3 magnitude brightness variation in Figure (\ref{march19}) must have another cause.

With equal long equatorial axes $2b$  and short axis, $2a$, the centripetal and gravitational accelerations are equal in magnitude at the period

\begin{equation}
P_c = \left(\frac{3\pi}{G\rho}\right)^{1/2} \left(\frac{b}{a}\right)^{1/2}.
\label{oblate}
\end{equation}

\noindent For modest $b/a$ = 1 to 2, consistent with the radar-inferred shapes of critically rotating asteroids (e.g.~Ostro et al.~2006, Busch et al.~2011, Naidu et al.~2015), and $\rho$ = 2700 kg m$^{-3}$, Equation (\ref{oblate}) gives $P_c$ = 2.0 to 2.8 hours for the critical period below which mass could be lost centripetally.    This again matches the ``spin barrier'' detected in the rotations of  asteroids (Carbognani 2017) and is substantially smaller than the limit on the period, $P \ge$ 10.8 hr.

\subsection{Eclipsing Binary}
\begin{figure}
\epsscale{.80}
%\plotone{March19.pdf}
\plotone{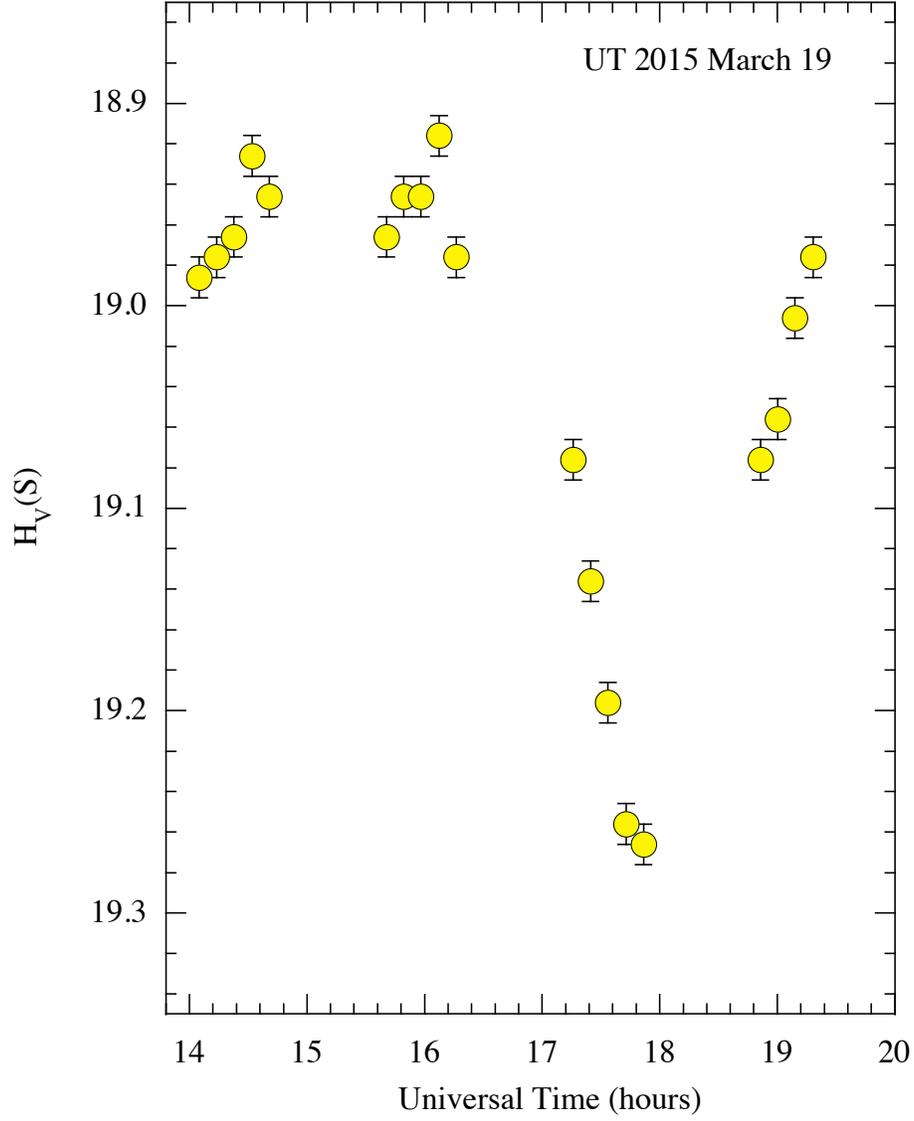}
\caption{Absolute V magnitude measured within a 0.2\arcsec~radius aperture as a function of time on 2015 March 19.  Plotted photometric error bars are $\pm$0.01 magnitudes. \label{march19}}
\end{figure}

The V-shaped drop in the brightness (at rates up to $\sim$0.4 magnitudes hour$^{-1}$)  shown in Figure (\ref{march19}) is reminiscent of the lightcurve of an eclipsing binary (e.g.~Lacerda and Jewitt 2007, Pravec et al.~2006, Scheirich and Pravec 2009, Pravec et al.~2016).   In this case, the V-shaped minimum would be caused by the transit or eclipse of a  secondary body in an unequal pair.  Representing both bodies as spheres having the same albedo and with radii $r_p$ (primary) and $r_s$ (secondary), we write

\begin{equation}
C_{max} = \pi \left(r_p^2 + r_s^2\right)
\label{cmax}
\end{equation}

\noindent at maximum light and

\begin{equation}
C_{min} = \pi r_p^2
\label{cmin}
\end{equation}

\noindent at minimum light, assuming a full transit.  With  $C_{max}$ = 0.13 km$^2$ and $C_{min}$ = 0.10 km$^2$, from \textsection (\ref{Activity}), we solve Equations (\ref{cmax}) and (\ref{cmin}) to find $r_p$ = 178 m and $r_s$ = 98 m for the radii of the two components.  The mass ratio of these bodies, if they are of the same density, is $m_s/m_p = (r_s/r_p)^3 \sim 1/6$. The observed duration of the supposed transit event, from first contact to last, is $\Delta T$ = 2 hours (Figure \ref{march19}).  From this, we compute the sky-plane velocity of the secondary as it crosses the face of the primary, using

\begin{equation}
 V = 2(r_p + r_s)/\Delta t,
 \end{equation}
 \noindent which gives $V$ = 0.077 m s$^{-1}$.  Assuming, for lack of evidence, that the orbit of the secondary is a circle, Kepler's law relates $V$ to the component separation, $r$, through
 
 \begin{equation}
 r = \frac{4\pi G \rho(r_p^3 + r_s^3)}{3 V^2}.
 \end{equation}
 
 \noindent Substituting, we find  $r$ = 840 m or, in units of the primary radius, $r/r_p =$ 4.2. The separation is a small fraction of the Hill radius (Equation \ref{Hill}), $r/R_H =$ 10$^{-2}$.  These properties are broadly similar to those of a large number of binary asteroids thought to have been formed by rotational fission (Pravec et al.~2006).

Detection of a transit would only be possible given a specific geometrical alignment of the binary system.  Is this alignment likely?   In order for a partial transit to be observed, the Earth must lie closer to the orbital plane of the secondary than  an angle $\tan(\theta) = \pm (r_p + r_s)/r$. We find $\theta =  \pm$0.36 radian ($\pm$21\degr). Given a random distribution of possible orbital poles, the likelihood of finding this alignment by chance is $\sim$62\%.  To observe a full transit (in which the silhouette of the secondary lies totally within that of the primary) requires $\tan(\theta) = \pm (r_p - r_s)/r$, which gives $\theta = \pm12\degr$.  The probability of finding even this more stringent  alignment by chance, given a random distribution of planes, is  still $\sim$40\%.  Hence, it is not statistically remarkable that we would detect mutual events given the small separation of this binary.

\begin{figure}
\epsscale{.90}
%\plotone{composite.pdf}
\plotone{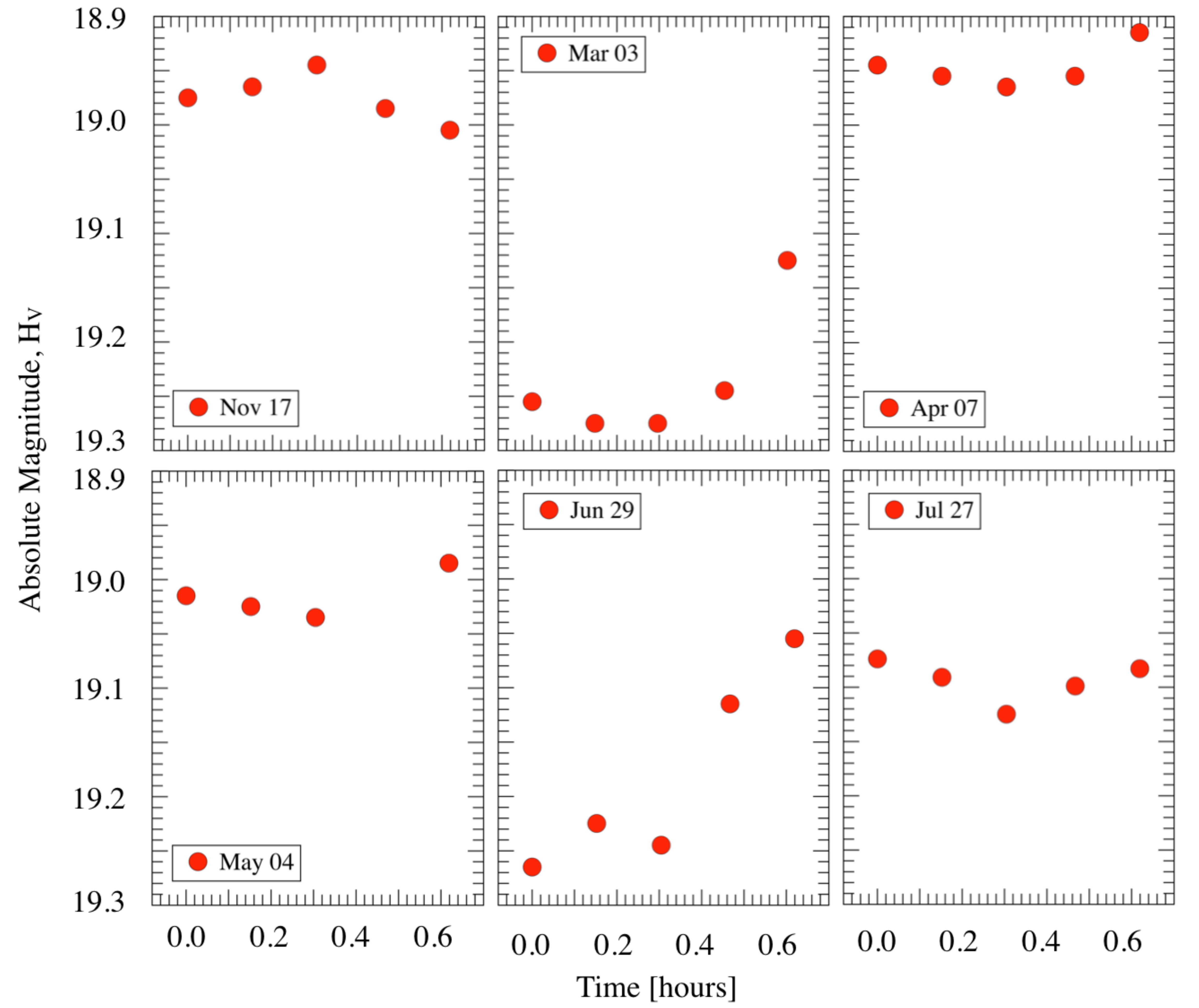}
\caption{Single orbit photometry plotted as a function of time in hours for six dates from 2014 November 17 to 2015 July 27.  The absolute magnitude is shown, computed according to Equation (\ref{abs}).  Formal errror bars, not shown, are comparable to the diameter of the symbols.
\label{composite}}
\end{figure}

The orbit period of such a system is $P_K = 2 \pi r /V$, or 6.8$\times$10$^4$ s (about 19 hours or 0.8 days), meaning that transits and mutual occultations are to be expected every $\sim$9.5 hours. There is therefore a high probability that we would detect one minimum in our 5.4 hour observational window.   Specifically, given that the HST observing windows are each $\delta t \sim$ 40 minutes (2/3 hour) in duration, the probability that a randomly sampled observing window will sample a deep lightcurve minimum is $4\delta t/P_K \sim$ 1/7.  In the 6 single-orbit observing windows shown in Figure (\ref{composite}), we should expect to find $\sim$1 deep lightcurve minimum.   The observations from March 03 definitely record the deep minimum, while observations from Jun 29 probably just missed it, apparently sampling the rise towards peak brightness.  

We conclude that mutual events are quite likely to be detected in a binary with $r/r_p$ = 4 and $P_K$ = 19 hours, given a randomly selected spin-pole and the temporal sampling of our HST observations.   However, while mutual events in a binary can explain the 0.3 magnitude dip in the lightcurve of 311P, they do not provide an explanation for the episodic mass loss indicated by the multiple tails ejected in 2013 (Jewitt et al.~2013, 2015b).  
 
Our suspicion is that the (``landsliding'') mechanism indeed drives the mass loss from one or both of the components, but that the rapid rotation required for the instability contributes little to the measured lightcurves because the unstable body has the rotationally symmetric (muffin-shaped) morphology described above.  For example, measurements from   individual HST orbits show short-term variations (Figure \ref{composite}) that are large compared to the errors of measurement (but small compared to the deep minimum recorded best on UT 2015 March 19; Figure \ref{march19}).  These short-term variations, which we have been unable to phase into a convincing lightcurve because of the strong aliasing in our widely-spaced data, could result from small deviations from rotational symmetry in a critically rotating ($P \sim$ 2 hour) muffin-shaped body.  

\subsection{Other Models}
Hainaut et al.~(2014) conjectured that dust emission from 311P might be due to friction at the contact point of a ``rubbing binary''.   However, it is not clear, in the context of a rubbing binary, why the interval between dust events should be measured in weeks and months (9 events spread over $\sim$250 days; Jewitt et al.~2013, 2015b) rather than being comparable to the orbital period, $P_K$ = 19 hours. Furthermore, in the absence of a continuous exciting force, and for any plausible values of the friction coefficient or the coefficient of restitution, we   expect that relative motion between binary components would  be quickly damped-out.   These concerns deserve exploration in a future, quantitative treatment of the physics of a binary rubble pile merger.

A final possibility is that 311P is in an excited rotational state, as a result of its small size and the long  damping time due to internal friction.   Either or both components of 311P could be non-principal axis rotators. Excited rotation is common in small asteroids because the damping times are long (Harris 1994) but typically much more data than we possess is required in order to identify the rotational state.  Therefore, we must leave this possibility as open.

\subsection{Formation}
A leading model for the formation of close asteroid binaries is by the rotational disruption of a weakly cohesive primary (e.g.~Walsh et al.~2012, Boldrin et al.~2016, Sanchez and Scheeres 2016, and see reviews by Margot et al.~2015, Walsh et al.~2015).  Spin-up is driven by radiation torques (the YORP effect) for which the characteristic time at 2 AU (estimated from published observations of asteroids in which YORP torque has been detected) is $\tau_{Y} \rm{(Myr)}\sim 2 (r_p/1~ \rm{km})^2$  (Jewitt et al.~2015a).  Beyond the $r_p^2$ size-dependence in this equation, the YORP timescale is very uncertain for any particular object because it is  dependent upon unknown details of the shape, spin vector and surface thermal properties.  Nevertheless, for an object as small as 311P ($r_p \sim$ 0.2 km gives $\tau_Y \sim$ 0.1 Myr) it is reasonable to expect that YORP torque has affected the spin.  
Binaries form when slowly-launched debris is captured and reassembled through dissipative collisions in orbit (e.g.~Jacobson and Scheeres 2011, Walsh et al.~2012).  The orbits of binaries so-formed are initially chaotic, with some reimpacting the primary and others escaping to infinity (forming asteroid pairs, Pravec et al.~2010, Polishook 2014).  Stabilization of the binaries can occur by the transfer of energy from orbital motions into the rotations of the components and by the ejection of mass (and energy and angular momentum) from the system.  Jacobson and Scheeres (2011) found that stable binaries formed only when the mass ratio is $m_s/m_p < 0.2$. This is consistent with our estimate $m_s/m_p \sim 0.15$ for 311P, but the modeling of disrupting systems is very sensitive to assumptions about the physical properties, most of which are unmeasured, and the agreement may not be significant.  A wide binary active asteroid (288P/(300163), with a separation $r/r_p \sim$ 100) has been discovered but was probably not formed by rotational fission and in-orbit reassembly alone (Agarwal et al.~2017).

\begin{figure}
\epsscale{.95}
%\plotone{Dp_vs_ratio.pdf}
\plotone{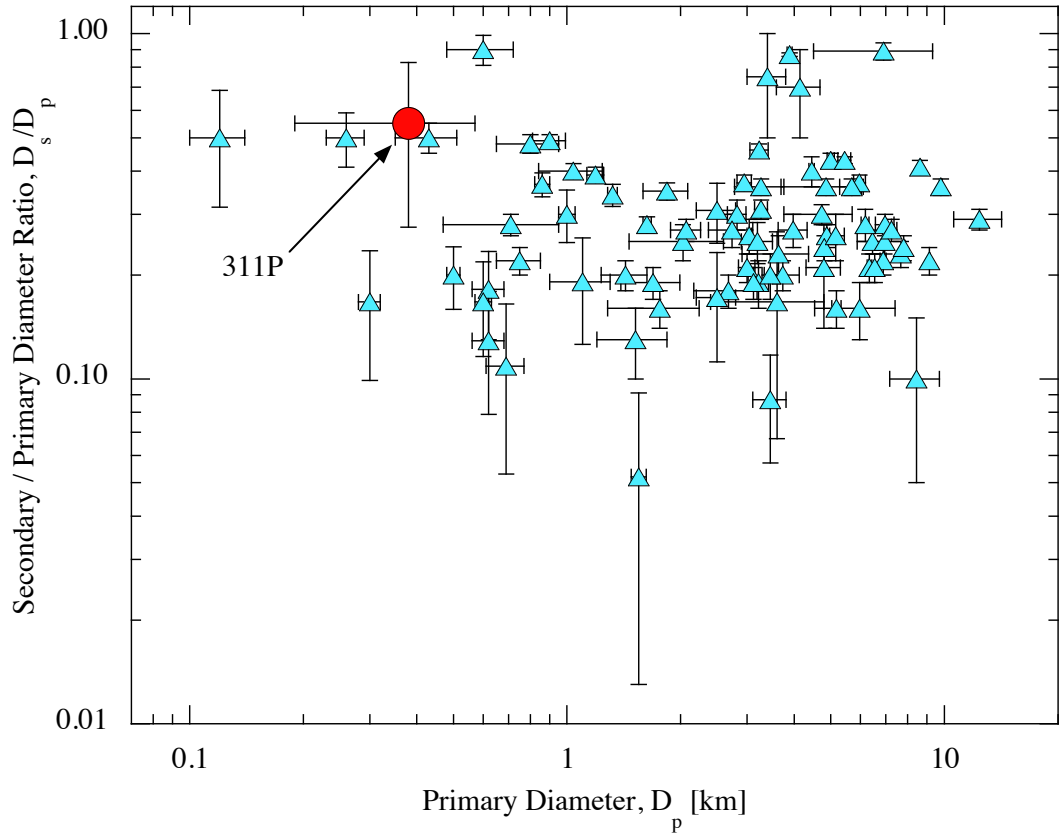}
\caption{Diameter of the primary component vs.~the ratio of secondary to primary component diameters. Asteroid data from the  compilation by Johnston (2016) are shown as blue-filled triangles while 311P is indicated by a red-filled circle.  \label{Dp_vs_ratio}}
\end{figure}

\begin{figure}
\epsscale{.95}
%\plotone{Dp_vs_orbit_ratio.pdf}
\plotone{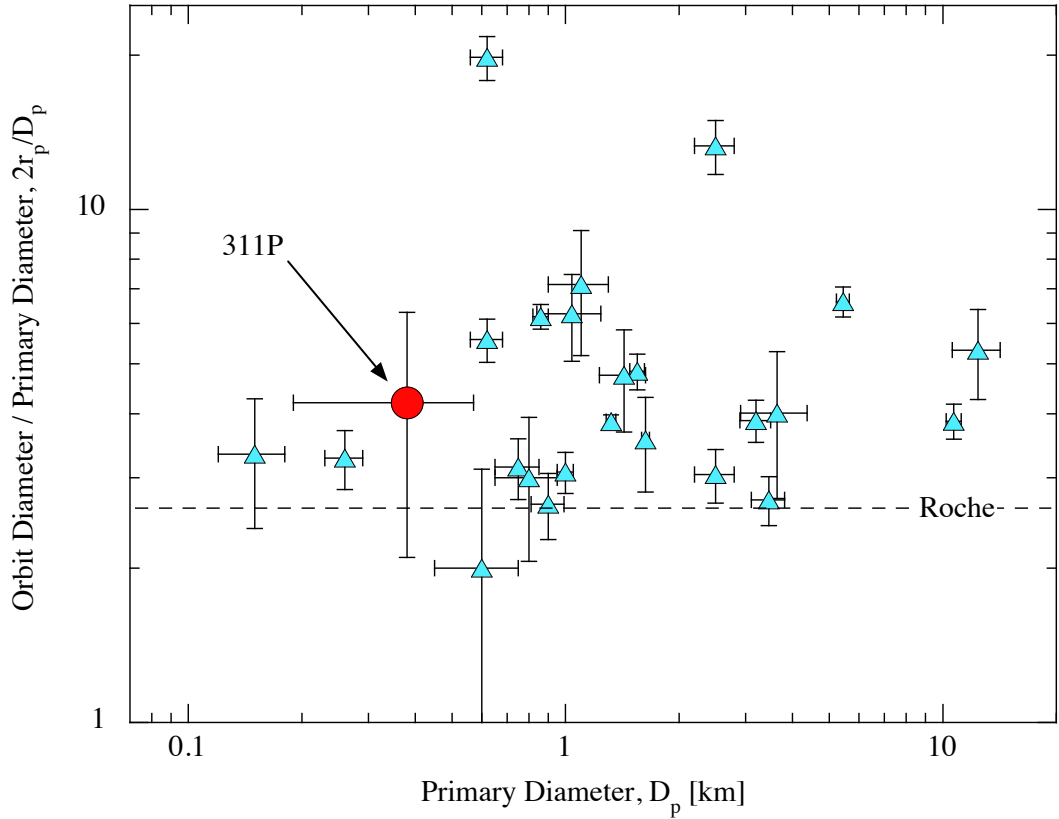}
\caption{Diameter of the primary component vs.~the ratio of the orbit diameter to the primary component diameter.  Asteroid data from the  compilation by Johnston (2016) are shown as blue-filled triangles while 311P is indicated by a red-filled circle. The dashed horizontal line shows the Roche limit for equal density spheres.  \label{Dp_vs_orbit_ratio}}
\end{figure}

We show  in Figure (\ref{Dp_vs_ratio}) a comparison between  the inferred secondary/primary object diameter ratio of 311P with the same ratio for  known binary asteroids having primary diameters $\le$ 20 km.  For the latter, we have used the compilation by Johnston (2016) and retained only those objects for which uncertainties on the primary and secondary object diameters are quoted.  The uncertainties on the parameters of 311P are unknown: we show  $\pm$50\% errors in diameter and diameter ratio, for illustrative purposes only.  While there are relatively few known sub-kilometer binary asteroids, the figure shows that the component diameter ratio of 311P is not atypical.  Similarly, in Figure (\ref{Dp_vs_orbit_ratio}) we compare 311P with the known binaries in the diameter vs.~orbital radius/primary radius plane, again taking data from Johnston (2016) and plotting only objects $\le$20 km in diameter for which uncertainties in both diameter and orbit radius/primary radius are quoted.  The dashed horizontal line at $r/r_p$ = 2.6 in Figure (\ref{Dp_vs_orbit_ratio}) shows the Roche limiting radius for equal density spheres.  It roughly coincides with the lower boundary of the measured binary separations, consistent with the expected nearly strengthless nature of secondaries formed by agglomeration in-orbit.  Again, 311P would be amongst the smallest  binary objects known (this is an artifact of its transient prominence caused by dust ejection) but it is not otherwise remarkable.  We conclude that the eclipsing binary interpretation of the 311P lightcurve is plausible.

Numerical simulations show that the evolution of rotationally disrupted aggregate bodies can be both complex and protracted (Boldrin et al.~2016).  Collisions and excitation of the spin of the secondary body can both dissipate orbital energy and so help to stabilize the asteroid as a close binary.  The timescales for system evolution can be surprisingly long compared to the orbit period of the system, measured in 100s and even 1000s of days.  If the detection of 311P in 2013 coincided with its initial breakup, then it seems entirely possible that the dynamical evolution may not yet be finished.  The multiple tails, for example, could result from equator-ward landsliding on a secondary being spun-up by mutual interactions, as well as from dust escaping from the unstable primary.  Radiation pressure-swept dust, kicked up from the primary by boulders infalling from temporary orbit, could also supply the episodic ejections.  Although no departing fragments larger than $\sim$10 m were detected in our search to date (\textsection \ref{search}), it is possible that such objects have yet to be launched and may be detected in future observations.  These intriguing possibilities make 311P an object highly deserving of continued study.

\clearpage

\section{Summary}
We present new observations of 311P taken to examine the properties of the nucleus in the context of suggestions that sporadic dust release from this body might be caused by rotational instability.

\begin{itemize}

\item We find a mean absolute magnitude $H_V$ = 19.14$\pm$0.02 which, with assumed geometric albedo $p_V$ = 0.29$\pm$0.09, gives a mean effective cross-section $C_e$ = 0.11$\pm$0.04 km$^2$. The radius of an equal-area circle is $r_e = (C_e/\pi)^{1/2}$ = 0.19$\pm$0.03 km.

\item No co-moving secondary bodies brighter than apparent magnitude $V$ = 27.8 exist beyond $\sim$1\arcsec~from the nucleus.  This corresponds to an effective radius of $<$11 m (again assuming $p_V$ = 0.29).

\item The main features of the lightcurve are a V-shaped, 0.3 magnitude deep minimum and a lightcurve period $P > $ 5.4 hour.  Interpreted as an eclipsing binary, we estimate a secondary/primary radius ratio $r_s/r_p$ = 0.55, a circular orbit separation to primary radius ratio $r/r_p \sim$  4 (about $10^{-2}$ of the Hill radius), and an orbit period $P_K \sim$ 0.8 days.  These parameters are consistent with those of numerous, small, near-Earth and main-belt asteroids that are thought to have formed through rotational instability and reaccretion.

\item The observed lightcurve reflects the orbital motion of the binary but provides no evidence for the rapid rotation inferred to drive centripetal dust loss from 311P.  However, rotational instability might still be the source of the ejected dust if one or both of the components of the binary has a nearly rotationally symmetric shape.  Such shapes are expected and observed in radar studies of rapidly rotating asteroids, where they are due to redistribution of particulate matter towards the rotational equator under centripetal accelerations.  

\end{itemize}

\acknowledgments

We thank the anonymous referee for comments.  Based in part on observations made with the NASA/ESA \emph{Hubble Space Telescope,} with data obtained via the Space Telescope Science Institute (STSCI).  Support for program 13866  was provided by NASA through a grant from STSCI, operated by AURA, Inc., under contract NAS 5-26555.  We thank Linda Dressel, Alison Vick and other members of the STScI ground system team for their expert help.  

{\it Facilities:}  \facility{HST (WFC3)}.




\clearpage

%% the file, uncomment the \plotone, \plottwo, and
%% If you need a layout that cannot be achieved with \plotone or
%% \plottwo, you can invoke the graphicx package directly with the
%% \includegraphics command or use \plotfiddle. For more information,
%% please see the tutorial on "Using Electronic Art with AASTeX" in the
%% documentation section at the AASTeX Web site, http://aastex.aas.org/
%%
%% The examples below also include sample markup for submission of
%% supplemental electronic materials. As always, be sure to check
%% the instructions to authors for the journal you are submitting to
%% for specific submissions guidelines as they vary from

%% This example uses \plotone to include an EPS file scaled to
%% 80% of its natural size with \epsscale. Its caption
%% has been written to indicate that additional figure parts will be
%% available in the electronic journal.

\clearpage

\clearpage

\clearpage

\begin{deluxetable}{lcclccccccr}
\tabletypesize{\scriptsize}
\rotate
\tablecaption{Observing Geometry 
\label{geometry}}
\tablewidth{0pt}
\tablehead{ \colhead{UT Date and Time} & Tel\tablenotemark{a} & DOY\tablenotemark{b}   & $\Delta T_p$\tablenotemark{c} & $\nu$\tablenotemark{d} & \colhead{$r_H$\tablenotemark{e}}  & \colhead{$\Delta$\tablenotemark{f}} & \colhead{$\alpha$\tablenotemark{g}}   & \colhead{$\theta_{\odot}$\tablenotemark{h}} &   \colhead{$\theta_{-v}$\tablenotemark{i}}  & \colhead{$\delta_{\oplus}$\tablenotemark{j}}   }
\startdata

2014 Nov 17 17:56 - 18:33 & HST & 686 & 215 & 78 & 2.111 & 2.280 & 25.7 & 291.6 & 295.6 & -1.5 \\
2015 Mar 03 14:30 - 15:06 & HST  &792 &  321 & 110 & 2.252 & 1.302 & 9.6 & 314.7 & 293.3 & 3.4 \\
2015 Mar 19 14:05 - 19:27 & HST  &808 & 337 & 115 & 2.272 & 1.283 & 3.8 & 21.3 & 294.1 & 3.8  \\
2015 Apr 07 16:19 - 16:57 & HST  &827 & 356 & 121 & 2.295 & 1.347 & 10.5 & 94.6 & 295.0 & 3.6 \\
2015 May 04 06:21 - 06:58 & HST  &854 & 383 & 128 & 2.323 & 1.558 & 19.9 & 107.3 & 295.4 & 2.7 \\
2015 Jun 29 21:03 - 21:40 & HST  &910 & 439 & 143 & 2.378 & 2.271 & 25.1 & 114.1 & 294.3 & 0.5 \\
2015 Jul 27 09:15 - 09:52 & HST  &938 & 467 & 150 & 2.399 & 2.616 & 22.8 & 115.1 & 292.9 & -0.8 \\
\enddata

%% Text for table notes should follow after the \enddata but before
%% the \end{deluxetable}. Make sure there is at least one \tablenotemark
%% in the table for each \tablenotetext.
\tablenotetext{a}{Telescope used: HST = 2.4 meter Hubble Space Telescope, Keck = 10 meter telescope}
\tablenotetext{b}{Day of Year, UT 2013 January 01 = 1}
\tablenotetext{c}{Number of days from perihelion (UT 2014-Apr-15.78 = DOY 259).}
\tablenotetext{d}{True anomaly, in degrees}
\tablenotetext{e}{Heliocentric distance, in AU}
\tablenotetext{f}{Geocentric distance, in AU}
\tablenotetext{g}{Phase angle, in degrees}
\tablenotetext{h}{Position angle of the projected anti-Solar direction, in degrees}
\tablenotetext{i}{Position angle of the projected negative heliocentric velocity vector, in degrees}
\tablenotetext{j}{Angle of Earth above the orbital plane, in degrees}

\end{deluxetable}

\clearpage

\begin{deluxetable}{lrcccc}
%\tabletypesize{\scriptsize}
\tablecaption{Average Nucleus Photometry 
\label{photometry}}
\tablewidth{0pt}
\tablehead{
\colhead{Date}      & DOY\tablenotemark{a} &  \colhead{$V_{0.2}$\tablenotemark{b}} & \colhead{$H_{0.2}(S)$\tablenotemark{c}} & \colhead{N\tablenotemark{d}}}
\startdata
2014 Nov 17     & 686 & 23.47$\pm$0.01 & 18.92$\pm$0.04 	& 5  \\
2015 Mar 03 	& 792 & 22.12$\pm$0.03 & 19.23$\pm$0.03	& 5 \\
2015 Mar 19 	& 808 & 21.67$\pm$0.03 &  19.03$\pm$0.03 	& 19 \\
2015 Apr 07 	& 827 & 21.98$\pm$0.01 &  18.95$\pm$0.01 	& 5 \\
2015 May 04 	& 854 & 22.71$\pm$0.01 &  19.04$\pm$0.01 	& 4 \\
2015 Jun 29 	& 910 & 23.85$\pm$0.04 &  19.17$\pm$0.04	& 5 \\
2015 Jul 27 	& 938 & 24.01$\pm$0.01 &  19.07$\pm$0.01 	& 5 \\

\enddata

%% Text for table notes should follow after the \enddata but before
%% the \end{deluxetable}. Make sure there is at least one \tablenotemark
%% in the table for each \tablenotetext.
\tablenotetext{a}{Day of Year, UT 2013 January 01 = 1}
\tablenotetext{b}{Apparent V magnitude within 5 pixel (0.2\arcsec) radius aperture.  Quoted uncertainty is the statistical error, only.}
\tablenotetext{c}{Absolute V magnitude, $H_{0.2}$ computed from $V_{0.2}$ assuming an S-type asteroid phase function and Equation (\ref{abs}).}
\tablenotetext{d}{Number of images used to compute the magnitudes on each date.}

\end{deluxetable}

\clearpage

\clearpage

\clearpage

\clearpage

\clearpage

\clearpage

\clearpage

\clearpage

\clearpage

%% The following command ends your manuscript. LaTeX will ignore any text

\end{document}